\documentclass[12pt]{iopart}

\usepackage{graphicx}
\usepackage{dcolumn}
\usepackage{bm}

\usepackage{graphicx}
\begin{document}

\title[WHY DON'T CLUMPS OF CIRRUS DUST GRAVITATIONALLY COLLAPSE?]
{Why don't clumps of cirrus dust gravitationally collapse?}

 \author{Rudolph E. Schild$^{1}$, Carl H. Gibson$^{2,4}$,  Theo M. Nieuwenhuizen$^{3}$, and N. Chandra Wickramasinghe$^{4}$}

\address{$^1$ Harvard-Smithsonian Center for Astrophysics, 60 Garden Street, Cambridge, MA02138,
USA\\}
 
\address{$^2$
Mech. and Aerospace. Eng. \& Scripps Inst. of Oceanography. Depts., UCSD, La Jolla, CA
92093-0411, USA\\ }

\address{$^{3}$Institute for Theoretical Physics, 
Science Park 904, P.O. Box 94485, Amsterdam, the Netherlands\\}

\address{$^4$Buckingham Centre for Astrobiology, The University of Buckingham, Buckingham MK18 1EG, UK\\ }

\ead{rschild@cfa.harvard.edu}

\begin{abstract}
We consider the Herschel-Planck infrared observations of presumed condensations of
interstellar material at a measured temperature of approximately 14 K (Juvela
et al., 2012), the triple point temperature of hydrogen.
The standard picture is 
challenged that the material is cirrus-like clouds of ceramic dust
responsible for Halo extinction of cosmological sources (Finkbeiner, Davis,
and Schlegel 1999). Why would such dust clouds not 
collapse gravitationally to a point on a gravitational free-fall time scale
of $10^8$ years? Why do the particles not collide and
stick together, as is fundamental to the theory of planet formation
(Blum 2004; Blum and Wurm, 2008) in pre-solar accretion discs? Evidence from
3.3 $\mu$m and UIB emissions as well as ERE (extended red emission) data
point to the dominance of PAH-type macromolecules for cirrus dust, but
such fractal dust will not spin in the manner of rigid grains 
(Draine \& Lazarian, 1998).  IRAS dust clouds examined by Herschel-Planck are easily 
understood as dark matter Proto-Globular-star-Cluster (PGC) clumps of primordial 
gas planets, as predicted by Gibson (1996) and observed by Schild (1996).
\end{abstract}

\maketitle

\section{Introduction}

When the IRAS infrared satellite (Soifer, Houck, and Neugebauer, 1987) discovered 
strong evidence of a new component of
infrared emission, it was
already understood that such cold emission was found in star forming
regions, and so was attributed to dust.  Hundreds of infrared
galaxies were soon discovered by IRAS emitting $\ge 95\%$ of their radiation
over 7 decades of luminosity from  $10^{6} L_{\odot}$ to $10^{13} L_{\odot}$. What is the source
of all this power?  Despite the low resolution of imaging in the DIRBE experiment, the emission was found to be thinly
extended across the entire Galaxy Halo in long filamentary structures
reminiscent of cirrus clouds. An improved representation of this can be
seen in Fig. 1 below,  the IRAS map of $T\approx$ 15K emission published as Fig. 1 of Juvela et al (2012). 
The review ``Cold Dark Clouds: The Initial Conditions for Star Formation'' by Bergin and Tafalla (2007) 
calls for the Herschel and Planck satellites launched in May 2009 to settle the many mysteries about
 the temperatures of the clouds of cirrus dust and their infrared luminous cold cores.

\begin{figure}[htbp] 
   \centering
\includegraphics[width=6in]{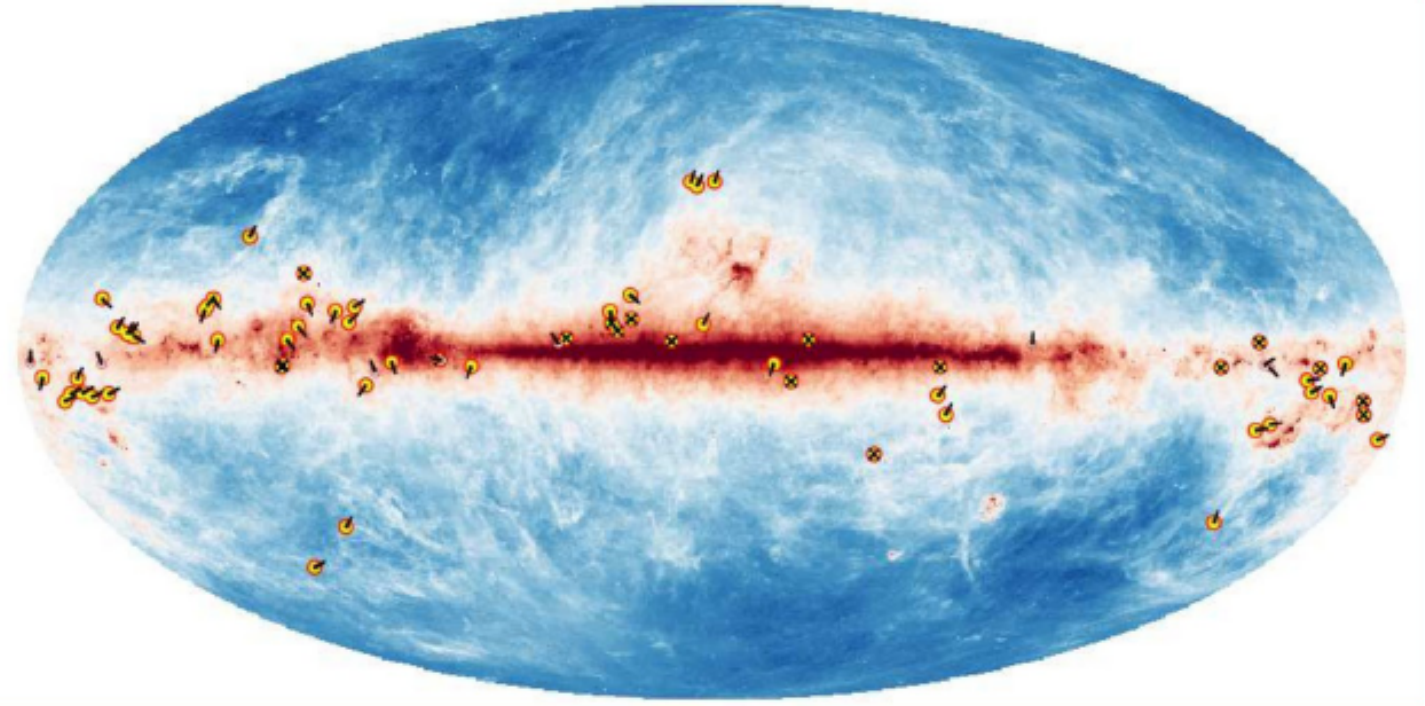} 
   \caption{IRAS 100 micron surface brightness map of the sky (Juvela et al. 2012 fig. 1, reproduced with permission of Astronomy and Astrophysics).}
   \label{fig:example}
\end{figure}

\begin{figure}[htbp] 
   \centering
   \includegraphics[width=15.7cm]{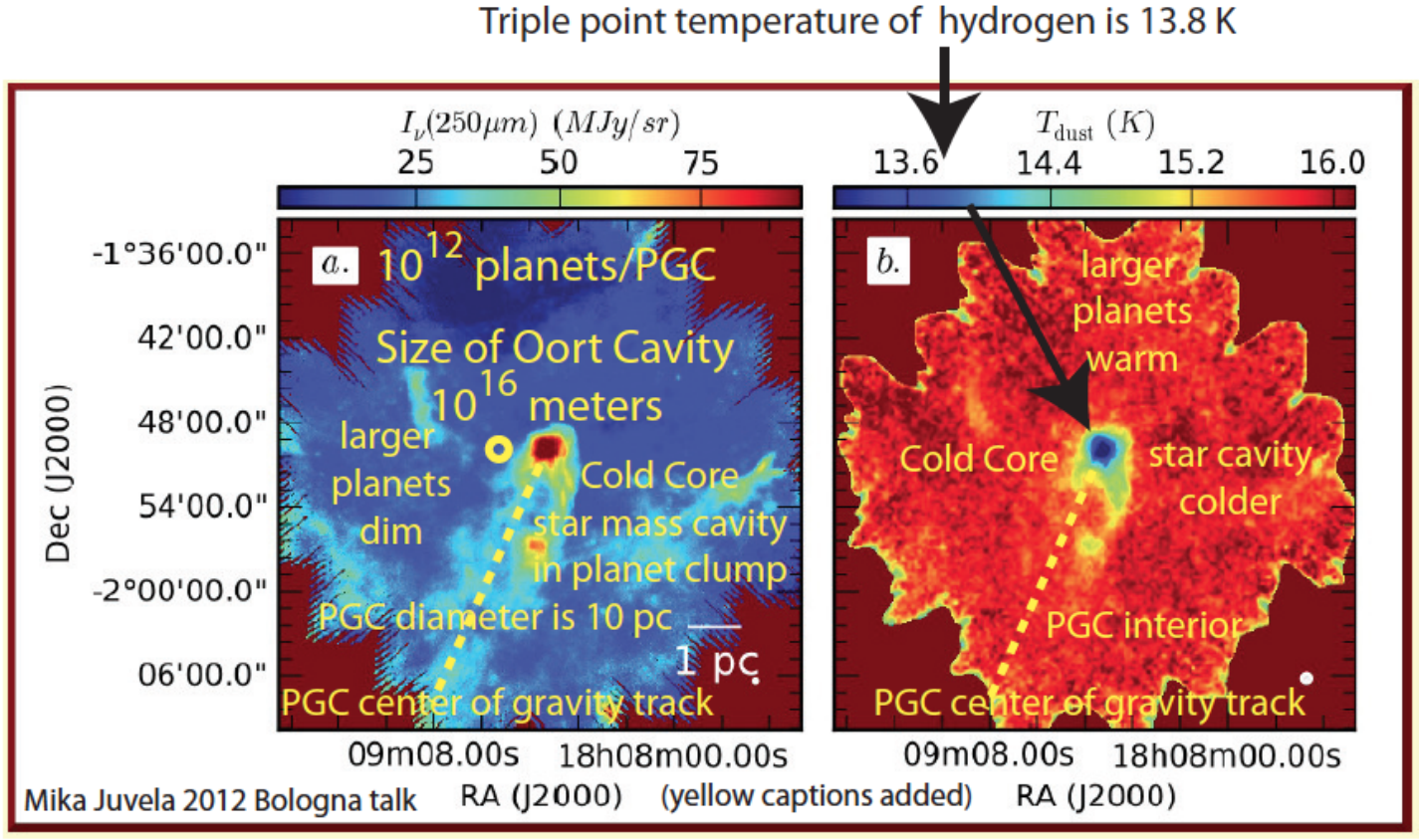} 
   \caption{Herschel-Planck infrared images of radiation and temperature 
within cold clouds, reported by M. Juvela   at the Planck 2012 Cold Cores Project meeting, Bologna and
reproduced with his kind permission.   Added to the original figure are our interpretations in yellow.
Image $a$ (left): An intensity map of what we understand as a PGC, to which we add the supposed track of a core. 
Image $b$ (right):  The radiation temperature of the Cold Core is brighter at the triple point of hydrogen, 13.8
   K, than the surroundings which are at higher temperature but lower surface brightness.  
   The HGD cosmology interpretation is that mergers of cold gas planets, rather than ceramic or PAH dust, 
   best explain the radiation and temperature patterns.}
   \label{fig:example}
\end{figure}

We have previously proposed (Gibson 1996, Schild 1996, Nieuwenhuizen et al 2011) that the observed  structures are in fact metastable
clumps of Jeans mass, fragmented in planet masses (PGC Proto-Globular-star-Clusters, also called Jeans Clusters) that formed primordially at the plasma to gas transition, 
and which comprise the missing baryonic dark matter of all galaxies. The clumping is initiated by Hydro-Gravitational-Dynamics (HGD) 
fragmentation of primordial plasma protogalaxies (Gibson 1996), and is
fractal (non-Gaussian) in statistical properties, and concentrically
structured, in clumps of clumps. An individual Jeans mass clump will
consist of a trillion Earth mass hydrogen spheres originally detected (Schild 1996) in
quasar microlensing. These will be in quasi-stable equilibrium and have
survived as coherent Halo structures for 13.7 billion years. When
disturbed they become the Giant Molecular Clouds that host star formation in an
accretional cascade from planet mass to stellar mass.


With much higher resolution, particularly of the Herschel telescope, the ``Cirrus Dust" is revealed to be a network of small condensations
carefully analyzed as regions within the disc and halo of our Galaxy  by Juvela et al. (2012). (See hereto Fig. 2 above.) The clumps were found and
studied at a number of distances and environments and have shown surprising consistency in their properties. From SPIRE photometry at wavelengths
250, 350, and 500 microns (combined with AKARI photometry at 90, 140, and 160 microns), 
temperatures estimated for 71 regions averaged (Juvela et al 2012, Table 2, column 6) 
about 14K, though estimates, where a background temperature was subtracted, were 1.5 K lower. A slightly different temperature was determined from
spectral energy distributions that introduce the longer-wavelength WMAP flux densities (Veneziani et al., 2010, 2012).

The understanding of the Herschel-Planck data in Figure 2 is confounding to astrophysics, because the situation revealed is opposite to the Planck radiation law that
ordinarily shows hotter radiating objects emitting at a higher intensity.  In contrast, Fig. 2 shows a pair of cold core sites of star formation 
radiating at a higher luminosity but at lower temperature, in contradiction to a straightforward application of the virial theorem.
We shall come back to this paradox in Section 2.

It is surprising that so little diversity was
found in the estimated temperatures. Because the radiation environment
would be expected to be very different for sources in the Galactic plane than for
sources (far) above the plane, it would be easier to understand the
temperature as being thermostatically controlled by the 13.8 K triple point phase
transition of Hydrogen, as proposed by Nieuwenhuizen et al. (2011). 
Indeed, before traversing the first order phase transition line from gas to solid which emerges from the triple point,  
the latent heat has to be radiated out.

At present it is accepted that the dust temperature is maintained
primarily by radiation from spinning dust particles (Hoyle and Wickramasinghe, 1970; Schlegel et al., 1997;
Draine \& Lazarian, 1998).  
An elongated rigid grain with average moment of inertia $I$ acquires a spin $\omega$  
by energy equipartition with a hot gas of temperature $T$ governed by the condition 
$I  {\omega ^{2}} /2= {3}k T/2$. 
For sufficiently small grains (small $I$) we may have rigid grains spinning at far infrared, microwave and 
radio frequencies. The ability of these grains, carrying a net electric charge and a non-zero dipole 
moment, to emit electromagnetic waves has been recognised for a few decades and is not in 
dispute (Hoyle and Wickramasinghe, 1970; Draine and Lazarian, 1998). However, in order to 
reproduce the observed flux data from SPIRE and AKARI photometry,  one has to require implausible 
and artificial constraints on any spinning grain model. Moreover, observations of 3.3 $\mu$m 
and other UIB emissions, as well as ERE (extended red emission) in cirrus clouds imply that 
cirrus dust is mostly composed of PAH macromolecules of undefined provenance 
(Szomuru and Guhathakurta, 1998; Smith et al, 2004; Wickramasinghe, 2010), with 
only minor contributions from silicate dust. Large PAH molecules with an inherent 
fractal structure will not behave like small mineral grains, so a rigid 
spinning grain 
model will not apply in this case.

A more plausible explanation of the observed infrared luminosity from HGD cosmology is that the merging of gas planets in PGC clumps releases large quantities of 
latent heat as the merging gas planets evaporate at the triple point temperature of the primordial hydrogen gas (14 K).

The morphologies of structures encountered in the observations is even more surprising.
Independently of where the clumps were located, whether at high or low
galactic latitude, the clumps were found in small clusters,  and
also in filamentary structures, though isolated clumps are more common at
high galactic latitudes (Juvela et al. 2012).

Although some of the clumps are quiescent at the low 14 K temperature, most
are apparently sites of star formation, as evidenced by the existence of
emission at mid-infrared wavelengths. The warmer emission co-exists with
the cold 14 K clumps.

About half of the fields studied show a filamentary structure which was not
predicted for ``dust cirrus clouds" before the HERSCHEL detection. 
The individual
filaments tend to be broken into individual knots of strong emission, also
not predicted before their observational discovery. The typical filament
width is 1/4 pc, measured as an infrared brightness FWHM, and sub-structure
within and along the extended filament structures is commonly found.  This
observed size closely matches the Oort cavity scale 
$L_{\rm Oort} \approx (M_{\odot} / \rho_0)^{1/3}=4\times10^{15}$ m from the  typical density of the PGC $\rho_0\approx 4 \times 10^{-17}$ kg m$^{-3}$.
This is the size of the cavity
produced in a PGC cloud of dark matter gas planets of density $ \rho_0$ when 
they merge to form a solar mass star, as shown in yellow font in Fig. 2a; in the Solar system $L_{\rm Oort}\sim0.5$ lyr
marks the transition from inner to outer Oort cloud.

In contrast with the view of the nature and distribution of interstellar
dust implied by the  HERSCHEL observations, the older view is 
very different.
The  prevailing view is that interstellar dust has condensed into
cores of planets orbiting stars in their pre-stellar discs, and that this
occurs in a process of grain growth, 
 collision and sticking (Blum and Wurm 2008), on a
time scale of approximately 1 million years, set by the observed ages of
young star-forming regions. Based upon laboratory data, it follows that the
porosity (or volume filling fraction) has a value of approximately 0.5, or
twice the density of liquid water. Experiments with crushing the ensembles
of growing dust particles of SiO$_2$ showed that filling factors in
the range 0.2 - 0.3 are measured, with corresponding porosities of
approximately 0.66 and densities of $1.5$ g cm$^{-3}$ (Blum, J. 2004). These
mechanical properties of interstellar cirrus dust are presumed to underlie
the process of planet formation.

However it is obvious that the HERSCHEL-observed 14 K temperature cannot
be maintained by cirrus dust whose temperature is maintained by spinning of
the dust particles. The Herschel image shows that the dust exists in clumps of
particles presumed to have been produced in collisional interaction, and
the presumed spin would have not credibly been maintained during the phase
of crushing by the self-gravity of the clumps. Thus we examine an
alternative to the identification of the radiating structures as dust.

\section{Alternative Explanation: Primordial gas planets contained in Jeans Clusters (PGCs)}

We offer an alternative to the cosmic dust condensation model for
producing the compact, centrally condensed regions of thermal emission
at 15 K discovered by Juvela et al. (2012) and others.  Our model
involves the hypothesis of gas spheres of planetary mass clumped in
proto-globular-star-cluster (PGC) (Jeans clusters) related to the
ordinary globular clusters observed in all galaxies.  As originally
proposed by Gibson (1996) from hydrodynamical considerations, these
structures would have formed in weak turbulence at the plasma-gas
transition at $z = 1100$ on scales of Jeans
mass, $5 \cdot 10^5 M_\odot$ (or perhaps only the $15,000 M_\odot$ of Nieuwenhuizen et al 2012)
and additionally at planetary mass, $10^{-6} M_\odot$. Such structure
was observed in quasar microlensing (Schild 1996) and described as ``rogue
planets" and as ``micro brown dwarfs'' by Nieuwenhuizen et al (2011).

Since the original discovery, the existence of rogue gas planets 
has been confirmed in 4
additional quasar lens systems having sufficient data to indicate a time
delay, but also with sufficient accuracy to reveal the rapid microlensing (see
Paraficz et al., 2005, for further references). Their wider detection and
properties have been further elucidated by Nieuwenhuizen et al. (2011), who introduce the term
``micro brown dwarfs'' for them.

The hydrodynamical theory that predicted their existence has further bearing upon the nature of structure that should be evident in the
low-$z$ universe. Following their formation as turbulent residual
perturbations they would have collapsed on a  timescale $1/\sqrt{G\rho_0}$ 
of approximately $6\,10^5$ years,  and thereafter cooled throughout the 13.7 Gyr
history of the universe, with their difficult cooling history lagging slightly behind the
lowering Cosmic Background Temperature of the Universe. An important
temperature in the physical chemistry of the universe is $T_{\rm tr}=13.8$ K, related to the triple point for
gas-liquid-solid phase transition in hydrogen gas, from which a first order gas-to-solid transition line emerges.
Lacking enough metals that would offer spectral lines, the difficulty to radiate out the latent heat would set the thermostat 
for hydrogen spheres with warm cores trying to radiate away their original
heat of gravitational collapse. The accumulated metals in their warm
cores would also produce heat by radioactive decay (Wickramasinghe
et al. 2010). Because the objects are observed to be cooling and
radiating at 14K, a temperature gradient in their atmospheres to drive
outward heat flow might leave a detectable signal of slightly higher
temperature, easily confused with core-shine.

Let us now return to Figure 2.
First, with the core temperature $T_{\rm core}$ lying at  $\sim13.6$ K, below $T_{\rm tr}$, we stress that this does 
not look like a run-away effect towards 2.7 K as may be anticipated naively.
Since the first-order transition line that emerges from the triple point goes towards smaller temperature for smaller pressures,
and the actual pressure likely lies below the triple point value $p_{\rm tr}=7.04$ kPa,  the  $T_{\rm core}$ temperature probably lies just above the transition line, as asserted by HGD.
Second, the virial theorem by itself does not allow the combination of cooling and radiating. Hence, to explain this property exposed in Fig. 2, 
energy has to be put into each primordial gas planet (rogue gas planet, micro brown dwarf). This can be achieved, in particular when there is a central mass,
by going into a more bound orbit, and using part of the gravitational energy gain for both radiation and cooling.
For a central mass of about one solar mass, a cooling by 2 K relates to the gravitational energy at a distance of 0.2 pc, about the radial scale of the central core in Fig. 2 a,b. 
Third, the higher intensity of the central zone in Fig. 2, despite its  lower temperature, exhibits, within this picture, an enhanced
concentration of these constituents. This enhancement of in-spiraling orbits 
clearly supports the proposition that star formation occurs by the merging of these primordial gas planets (Gibson 1996).

 The picture can also be tested in our immediate neighborhood. 
 The Local Leo Cold Cloud is  a large ($\sim5$ deg), cold (15-20 K), nearby (10-20 parsec) H cloud (Peek et al 2011).
 Like the ill-understood  Lyman-alpha clouds (Rauch 1998), 
 it can be long-lived if it contains many unresolved rogue gas planets or micro-brown dwarfs, that bind the gas gravitationally,
 thus explaining why the gas can have the observed large pressure discussed in (Meyer et al 2012).
 The observed velocity differences inside the Local Leo Cold Cloud are explained by the local winds due to mostly  random motion of these gas planets
 that sweep the gas with them.

\section{Comparison of Predicted Structure to Sub-mm Observations}


The Juvela et al. (2012) analysis of the clumping of matter seen at the
wavelength peak of 15K emission gives an important perspective on the
evolution of primordial structure and its evolution to presently observed
(local umiverse) structure. They report structure dominated by knots of 0.32
pc diameter in clumps and filaments of characteristic diameter 6.1 pc. Structures like these
must have evolved into the globular clusters seen today with a mean
diameter of 6.6  pc. (All dimensions are expressed as FWHM and the knot and
clump diameters are averages from Juvela et al tables 2 and 3. The quoted
mean diameter for globular clusters is from the Harris 1996 compendium). Thus we
conclude that the clumps of knots seen today as 15K emission are precursors of today's globular clusters.  

In an accompanying paper Nieuwenhuizen et al (2012) point out that several  thousands  of  cold $\sim15$ K clouds can be discriminated 
in the Halo towards the Magellanic Clouds, the properties and distribution of which is consistent with them being PGCs (Jeans clusters)
created by the Jeans instability and HGD fragmentation into Earth mass-scale gas clouds,
that comprise all the missing dynamical mass of the Galaxy.

\section {Acknowledgement}

We acknowledge and thank Dr. W. E. Harris for maintaining the Globular Cluster
Data Base, available at:
http:/$\!$/physwww.mcmaster.ca/$\%$/WEHarris.html.
We have used this data base to determine the mean globular cluster diameter.
We thank Professor Mika Juvela for his excellent comments and advice about this paper.

\end{document}